\documentclass[12pt,preprint]{aastex}
\usepackage{times,amsmath}

\def\chandra    {{\em Chandra}\/}

\begin{document}

\title{Kinetic SZ effect and CMB polarization from subsonic bulk motions of 
dense gas clouds in galaxy cluster cores}

\author{J.M. Diego$^{1,4}$, P. Mazzotta$^{2,3}$, J. Silk$^1$}

\begin{abstract}

Recent \chandra ~ observations  have revealed 
the presence of cold fronts in many clusters of galaxies.
The cold fronts are believed to be produced by the bulk motions of 
massive,  dense, cold gas clouds with respect to the hotter, more 
rarefied ambient gas at velocities that can be as high as the speed of sound.
This phenomenon may produce a significant contamination of both the kinetic SZ 
effect and the CMB polarization pattern observed in the 
direction of a cluster. We estimate the contributions 
to the kinetic Sunyaev-Zel'dovich (SZ) 
effect and  to the CMB polarization toward galaxy clusters produced by
the bulk motions of the gas in the inner parts of galaxy clusters. 
The observed cold fronts probe the absolute velocities of the gas motion
while the induced polarization and  the kinetic SZ effect
probe the transverse and the radial components, respectively.
We show that these signals may  be easily detected with  sensitive future
experiments, opening an exciting  new window for studies of galaxy cluster 
internal dynamics, and eventually facilitating reconstruction of the
intrinsic cluster polarization of the CMB and the associated measure
of the local CMB quadrupole.
\end{abstract}


\altaffiltext{1}{Department of Physics (Astrophysics), University of Oxford,
Denys Wilkinson Building, 1 Keble Road, Oxford OX1 3RH, United Kingdom}
\altaffiltext{2}{University of Durham, Department of Physics, South Road, Durham DH1 3LE, UK.}
\altaffiltext{3}{Harvard-Smithsonian Center for Astrophysics, 60 Garden st., Cambridge MA02138, USA.}
\altaffiltext{4}{email: jmdr@astro.ox.ac.uk}

\section{Introduction}\label{sect_intro}
Recent observations with \chandra ~ have unveiled the presence of cold fronts in many 
galaxy clusters (Markevitch et. al. 2000; Vikhlinin, Markevitch, and  Murray 2001,
Mazzotta et. al 2001; Markevitch et al. 2002). 
The cold front is a discontinuity in the gas density and temperature characterized 
by the fact that the denser region  is significantly colder than the rarefied component 
(thereby producing a pressure jump which is smaller than 
either the density or the temperature jump).
Although the first cold fronts were discovered in merging clusters, 
\chandra ~ later revealed the presence of cold fronts 
in many apparently relaxed clusters (e.g. Mazzotta et al. 2001; Markevitch et al. 2001; 
Mazzotta, Edge, and Markevitch 2003).
Furthermore, a preliminary analysis shows that 2/3 of a 
X-ray flux limited sample of clusters of galaxies host cold fronts 
(Markevitch, Vikhlinin, Forman 2002). 
Since  due to projection effects,  only $\approx 1/3$ of all cold fronts can be 
observed in  X-rays (Mazzotta et al. 2001), this strongly indicates 
that essentially all  clusters of galaxies (including the most relaxed ones) 
host at least one cold front.

The most likely explanation for the presence of a cold front is that
the cluster hosts  a dense cold gas cloud which is moving with
respect to the hotter and more rarefied ambient gas (Vikhlinin, et al. 2001).
So far it is not completely clear why so many clusters 
(especially rounder, relaxed clusters) host moving gas clouds.
One possibility is that the gas cloud in the central region is oscillating 
(sloshing) around the minimum of the cluster potential well 
(see. e.g. Markevitch et al. 2001; Markevitch et al. 2002; 
Mazzotta et al. 2003). This bulk gas motion may either 
be the result of a recent merger or due to dynamic feedback from  the
central (not necessarily currently active) AGN.   

An important outcome of the study of cold fronts
is that  the gas thermal pressure at the front 
can be used  to estimate the speed of the moving 
gas cloud (Vikhlinin et al. 2001).
Preliminary studies have already revealed that these moving gas clouds
have a wide range of sizes and velocities. 
More specifically: i) the size of the moving gas cloud 
can be as large as twice the cluster core radius; ii)
the velocity is generally  subsonic or at most transonic  
in rounder systems (e.g A1795, Markevitch et al. 2001; 
RXJ~1720+26, Mazzotta et al. 2001; 2A~0335+096, Mazzotta et al. 2003) 
and can be even supersonic in merger systems 
(e.g. A3667 Vikhlinin, et al. 2001; 1E~0657-56,  Markevitch et al. 2002).  
 It is interesting that the bulk motions, combined with the higher 
densities, of these gas clouds may induce a kinetic SZ effect as well
as the polarization of the CMB. 
In this Letter, we study the contributions to the kinetic 
Sunyaev-Zel'dovich (SZ) 
effect and  to the CMB polarization toward galaxy clusters produced by
the bulk motions of these moving gas clouds in the inner parts of 
galaxy clusters. 


\ \

\ \

\section{Kinetic SZ effect and polarization signals}
The kinetic SZ effect is a distortion of the CMB spectrum due to the 
peculiar motion of the cluster with respect to the Hubble flow. This distortion 
is given by the expression,
\begin{equation}
\frac{\Delta T}{T} = - \frac{\rm{v}_r}{c}\tau \equiv - \frac{\rm{v}_r}{c}\int \sigma_T n dl  
\label{eqn_kinetic}
\end{equation}
where $\rm{v}_r$ is the radial component of the velocity of the cluster, 
$\tau$  the optical depth of the cluster, 
$\sigma _T$  the Thomson cross-section, and $n$ the cluster gas density. 
This distortion is the same at all frequencies. 
The same CMB photons which undergo Compton scattering will be polarized after the 
scattering. 
The net polarized signal is 
proportional to the quadrupole of the incoming radiation (Rees 1968, Polnarev 1985). 
If the cluster is moving with a given velocity with respect to the Hubble flow, 
there will be an induced quadrupolar signal (Zel'dovich \& Sunyaev 1980). 
\begin{equation}
P = \frac{\sqrt{Q^2 + U^2 + V^2}}{I_{tot}} = 0.1 f(x) \left( \frac{\rm{v}_t}{c} \right)^2 \tau
\label{eqn_polarization}
\end{equation}
where $I_{tot}$ is the total intensity of the radiation, $Q, U, V$ are the other  
Stokes parameters ($V=0$ for Thomson scattering), 
$\rm{v}_t$ is the tangential component of the peculiar velocity 
of the cluster (cluster velocity projected on the sky plane) and $f(x)$ is the 
frequency dependence of the polarized signal (Sazanov \& Sunyaev 1999),
\begin{equation}
f(x) = \frac{x^2 e^x(e^x + 1)}{2(e^x - 1)^2}
\end{equation}
with $x  \approx \nu (GHz)/56.8$.
The resulting polarization plane is perpendicular to the velocity vector. 
As noted in Sazanov \& Sunyaev (1999), there are other sources of CMB polarization 
in galaxy clusters. 
One of the most important is due to the intrinsic quadrupole of the CMB. 
For the COBE quadrupole, they estimated that the maximum signal should 
be $\approx 5 \tau \ \mu K$ (Rayleigh-Jeans) which for typical cluster 
optical depths gives $\approx 10 \ nK$. 
As we will see later, this value is comparable to the contribution of 
the cluster peculiar velocity plus the bulk motion of the central region. 
There are other mechanisms producing polarisation in the cluster but they have been  
discussed elsewhere (finite optical depth: Sazanov \& Sunyaev 1999; 
cluster rotation: Chluba \& Mannheim 2002). 

\section{Cold fronts: A toy model}
Studies of cold fronts in clusters of galaxies have shown
that many (if not all) clusters of galaxies host dense, cold gas 
clouds that move with respect to the cluster itself. 
These gas bulk motions may affect both the intrinsic 
kinetic SZ and the CMB polarization signal induced 
by the cluster.  
To predict the strength of the signal 
we  make very simple assumptions using a  toy model 
to describe the dynamics  of the  cold fronts observed in the X-rays.
The basic elements of our toy model are the cluster within which is embedded 
a dense gas cloud.
To fully define the model, we only need to establish the 
{\it geometry}  and its {\it dynamics}. 

We define the geometry by assuming that both the cluster and the 
moving gas cloud have spherical symmetry. 
To maintain the spherical symmetry of the problem,
we also assume that at the time of observation, the center of the moving
gas cloud and the cluster center coincide.
The displacement of the moving cloud with respect to the cluster center is
discussed later. In accordance with observations, we assume that 
the cluster gas density profile is described
by  a $\beta$-model with $\beta=2/3$ (Jones \& Forman, 1999)
that we truncate at the virial radius $r_{v}$:
$n(r) = n_0/(1 + (r/r_c)^2)$.
In the previous equation, $r_c$ is the cluster core radius
and $n_0$ is the central gas density that we obtain after 
fixing the total mass (and baryon fraction) of the cluster.
For convenience, we assume that the virial radius and the core radius
of the cluster are related by  $p=r_v/r_c $, with $p=10$. 
In the same way we  parameterize the real size of the moving 
gas cloud $r_T$ as $s=r_T/r_c$ and we take the same central density $n_0$ 
(although other values could be considered).

The dynamics of the model  are defined by  assessing 
the velocity of the gas at each point of the plasma. 
In our toy model, we simply assume  that the cluster is moving with its
own peculiar velocity $\vec{v}_1$ while the central gas cloud
has a bulk motion $\vec{v}_2$  with respect to the 
cluster itself. 
It is convenient to parametrize the gas cloud velocity  
as  $\vec{v}_2= M v_s\hat{n}$, where $M$ is the Mach number,  
$v_s$ is the cluster speed of sound, and $\hat{n}$ the gas cloud 
direction of motion. 
It is important to note  that while $\vec{v}_1$ 
is directly referred to the Hubble flow, $\vec{v}_2$ is given 
in the framework in which the cluster is at rest. 
Therefore, to obtain  the final velocity of the moving gas cloud
 with respect to the 
Hubble flow, we need to add  $\vec{v}_2$ to $\vec{v}_1$. 

With this simplistic description it is easy to reduce the problem to 
a two-phase plasma with a discontinuity in the velocity. 
From Eqs.~\ref{eqn_kinetic} and \ref{eqn_polarization}, we see that
the signal (kinetic SZ and/or polarization) at any  point of the 
cluster  can easily be computed  as:
$S = (\tau_{cluster}-\tau_{cloud})g(\vec{v}_1)
+ \tau_{cloud} g(\vec{v}_1+\vec{v}_2)$ 
where $\tau_{cluster}$ and $\tau_{cloud}$ are the
optical  depths of the entire cluster and the moving gas cloud respectively
and we introduce the new function $g(\vec{v})$ which is
$g(\vec{v})\propto v_r$ and $g(\vec{v})\propto (v_t^2)$ 
for the kinetic SZ effect and CMB polarization, respectively.  
We can also define the percentile variation 
\begin{equation}
\Delta S_0/S_0\equiv (S-S_0)/S_0 = 
\frac{\tau_{cloud}}{\tau_{cluster}}\frac{\Delta g(\vec v_{1},\vec v_{2})}{g(\vec v_{1})}
\label{eq:signalpercentile}
\end{equation}
where $S_0\equiv\tau_{cluster}g(\vec{v_1})$.
Using our toy model we can calculate $\tau_{cloud}$ and 
$\tau_{cluster}$ as a  function of the angular 
distance  $\theta$ from the cluster center.
The gas cloud optical depth is given by:
\begin{equation}
\tau_{cloud}\equiv \tau(\theta,s) = \frac{2 n_o r_c \sigma _T}{\sqrt{1 + 
(\theta/\theta _c)^2}}\tan^{-1} \left[ \sqrt{\frac{(s)^2 - (\theta/\theta _c)^2}{1 + 
(\theta/\theta _c)^2}}\right],
\label{eqn_tau}
\end{equation}
where $\sigma_T$ is the Thomson cross-section and $\theta_c$ is the 
angular dimension of the core radius. 
In the same way,  $\tau_{cluster}$ can be obtained from the previous expression
by substituting $s=p$.
%
We can illustrate the relative importance of the gas cloud in the center 
of the cluster by calculating the ratio: 
$\tau_{cluster}/\tau_{cloud} \approx 2$
This is particularly important as it tells us that,
despite  the small {size of the cloud ($\approx 2.5\%$ of the total mass), 
the optical depth of the central  moving gas cloud  in the cluster center
is  comparable to that of the remaining part of 
the cluster ($\tau_{cloud}\approx \tau_{cluster}- \tau_{cloud}$). 
The above ratio can change for different models but for 
reasonable models it is close to 2. 
We can see that, in this case, the  percentile variation of the signal 
in the innermost cluster region is approximately
half of the percentile variation of the function $g$.   
As an example,  let us consider  $T=9\,{\rm keV}$. 
Typical velocity values for such a  cluster are
$|\vec{v}_1| = 300 {\rm km} \, {\rm s}^{-1}$ and 
$|\vec{v}_2| =  1500  M \, {\rm km}\,{\rm s}^{-1}$. 
For convenience, we  assume that i) 
the gas cloud moves at the sound speed ($M=1$);
ii) the direction of motion of the cluster and the central cloud form 
 angles of 45$^{\rm o}$ with respect to the line of sight 
(in such a way $v_r$=$v_t$ and both the cluster and the moving cloud 
contribute to the kinetic SZ effect and the CMB polarization);
iii) the direction of motion of the cluster is in the y-z plane
while the direction of motion of the the central cloud is in the 
x-z plane (the z-axis is aligned with the line of sight). 
In this simple situation for the kinetic SZ effect, we find that 
$\Delta g(\vec{v}_1,\vec{v}_2)/g(\vec{v}_1)= 5$. 
This means  that because of the moving  dense gas cloud, 
in the  innermost cluster region, 
the  SZ effect induced by this cluster is $\approx 250\%$ higher 
than in the  ``classical'' case in which such a motion is not taken 
into account. If we calculate the polarization, the effect is even 
more dramatic since the signal is proportional to the 
square of the velocity. In particular 
$\Delta g(\vec{v}_1,\vec{v}_2)/g(\vec{v}_1)= 26$, which means that the 
polarization effect is $\approx 1300\%$  higher!
Interestingly, due to the dependence on  the square of the velocity, 
the polarization signal is 
insensitive to the sign of the velocity. 
Two regions having velocities with opposite sign 
(but the same magnitude) 
will contribute similarly  to the polarization.
Therefore, a cancellation of the polarization signal through 
averaging the signal over the cluster is less likely than in the case of 
the kinetic SZ effect (unless the velocities are orthogonal, in which case 
they produce a cancellation in the polarization). \\
To better illustrate the contribution of the gas cloud to the SZ effect
and CMB polarization,   we show in  Fig.~\ref{fig1} 
the maps of these signals induced by the toy model described above.
In these maps, the gas cloud is displaced from the cluster center
by 60 kpc in both x and z directions.
Such a displacement is seen in many round clusters with cold fronts
(e.g. A1795, 2A 0335+096, RXJ1720+26)  consistently with the fact that 
the central cluster region is moving with respect to the cluster.
Clearly, the displacement of the moving region  introduces an 
asymmetry into the previous problem making it rather  more complicated
to interpret. 
The figure shows an evident enhancement of both the kinetic SZ effect 
and CMB polarization over  the entire region corresponding to the 
size of the moving cloud.
For clarity in the  figure, we also show the kinetic SZ effect along 
a linear scan crossing the center of the cluster of the toy model 
(continuous line) and the same signal corresponding to a model 
with no moving gas cloud (dashed line).
We notice that i) the peak of the signals are shifted which is a simple
consequence of the dispacement of the moving gas cloud; ii)
the SZ signal of our toy model
is $\approx 3$ times higher than the ``classical'' signal 
for most of the cluster region corresponding to the 
size of the moving cloud.  
For the CMB polarization we find that the maximum amplitude 
signal for the toy model is 340 nK 
(at 300 GHz).
In the Rayleigh-Jeans regime, this amplitude drops to 23 nK.
This number is comparable to  (or even higher than)  the estimated 
amplitude of the polarization from the local CMB quadrupole 
(Sazanov \& Sunyaev 1999). 
The maximum polarization signal can be compared with the expected maximum 
signal in the classical case (no moving cloud), 25 nK (300 GHz).\\ 
To conclude the description of our toy model,
in  Fig.~\ref{fig2} we show the SZ effect and CMB 
polarization maps in
 the case in which the direction of motion
of the cluster is opposite to that of the gas cloud.
We notice that with respect to the other case considered,
the amplitude of the polarization is smaller. 
This is simply due to the fact that the velocity of the cluster 
partially compensates the velocity of the moving clouds.
Much more interesting, however, is the signature of the kinetic SZ effect. 
It shows a unique feature: a positive peak in the middle 
of the negative well. 
This feature can be used in high resolution, high 
sensitivity experiments to unequivocally identify bulk gas motions
inside clusters of galaxies.
\begin{figure}
\plotone{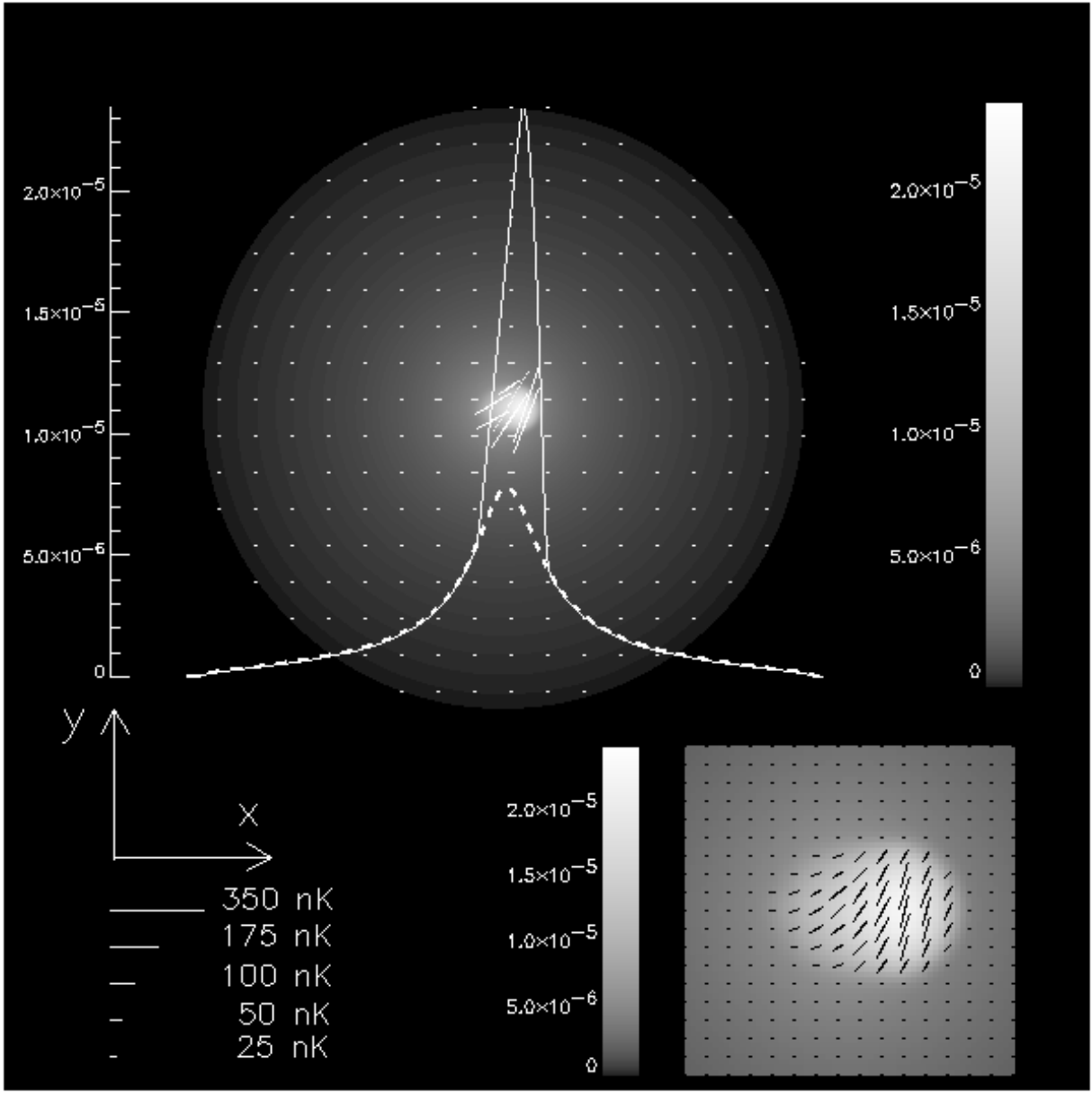}
\caption{
Kinetic SZ effect ($\Delta T/T$, gray scale) plus polarization 
signal (short lines) for the case where 
the central core and the cluster have positive velocity along the line of 
sight. 
In both cases we have assumed that the velocity vectors form 45$^{\circ}$ 
with the line of sight so the cluster and central region contribute to both 
polarization and kinetic SZ effect. 
Moreover, we assume that the cluster is moving in the 
plane y-z and the central region in the plane x-z. 
The z-axis is pointing to us. 
The solid line plot shows the kinetic component along a linear scan crossing 
the center of the cluster.
The dashed line is the kinetic SZ effect for a model with no moving gas cloud.
The maximum polarization signal 
is 340 nK (300 GHz). The lower right corner shows a zoomed region around the 
center of the cluster.
The radius of the cluster corresponds to a 
physical size of $1.3h^{-1}$~Mpc.
}\label{fig1}
\end{figure}
\section{Discussion and Conclusions}
In the previous section, we showed that the bulk motion of a
dense gas cloud in the cluster center may 
significantly affect both the  kinematic SZ effect 
and the CMB polarization signal induced by clusters of galaxies. 
On the one hand, this effect  represents a strong source of noise for
all  experiments that require
a precise measurement of the  SZ effect and cluster-induced 
CMB polarization. On the other hand, it provides a new tool
to better investigate the cluster structure. 
In the following we discuss these two aspects.
\subsection{Bulk gas motion as a possible source of noise} 
Peculiar velocities of galaxy clusters 
are an important cosmological 
tool since they can be connected with the velocity field of 
density perturbations (see e.g Chiu et al. 1998, Colberg et al. 2000, 
Sheth \& Diaferio 2001).
The kinematic SZ effect is thought to be one of the most promising
mechanisms for  measuring these peculiar velocities 
(e.g Haehnelt \& Tegmark 1996, Holzapfel et al. 1997, 
Aghanim et al. 2001).\\
In a similar way, the measurement of the CMB quadrupole in 
different areas of the universe may be the only 
way to solve such  questions as   whether or not we live 
in a universe with a particularly low quadrupole or we live in a 
universe with a peculiar geometry (Kamionkowski and Loeb 1997) 
or we are seeing   a possible signature of new physics (e.g Efstathiou 2003, 
Cline et al. 2003, Feng \& Zhang 2003).
The measurement of the polarization signal 
(due to the CMB quadrupole) 
in galaxy clusters can also be used as a probe 
of the dark energy which would boost the 
CMB quadrupole by the integrated Sachs-Wolfe effect 
(Cooray \& Baumann 2003). \\

The bulk gas motion can represent a serious issue 
for all the proposed cosmological studies devoted
to measuring the peculiar velocities of galaxy clusters 
including  kinetic SZ and/or local CMB quadrupole 
measurements from polarization. 
In the case of the kinetic SZ, this problem is solved if one assumes that 
the total signal is still proportional to the total average momentum of the cluster.
These studies can be also affected 
by other secondary effects such as the 
kinetic SZ effect and polarization associated with  cluster rotation 
(Chluba \& Mannheim 2002). \\
A way to circumvent this 
problem is by using data from high-resolution,
 high-sensitivity experiments which can 
disentangle the central from the external signal or by means of
 multifrequency observations 
which could disentangle the kinetic polarization term from the 
contribution due to the scattering of the CMB quadrupole. 
An estimate of the cluster velocity after averaging over the 
kinetic SZ map can 
produce unbiased estimates of the peculiar cluster velocity 
with a dispersion of 
50-100 km s$^{-1}$ when the velocities are averaged over an 
aperture at least as large 
as the virial radius (Nagai et al. 2003). 
Also, averaging the polarization measurements toward galaxy 
clusters in a region 
of the sky would help to reduce the contribution from the kinetic 
polarization term 
(Cooray \& Baumann 2003). However, the number of clusters which
 need to be {\it averaged} 
will depend on the dispersion of the kinetic polarization term. 
The effect studied in this 
Letter will increase this dispersion.\\
\begin{figure}
\plotone{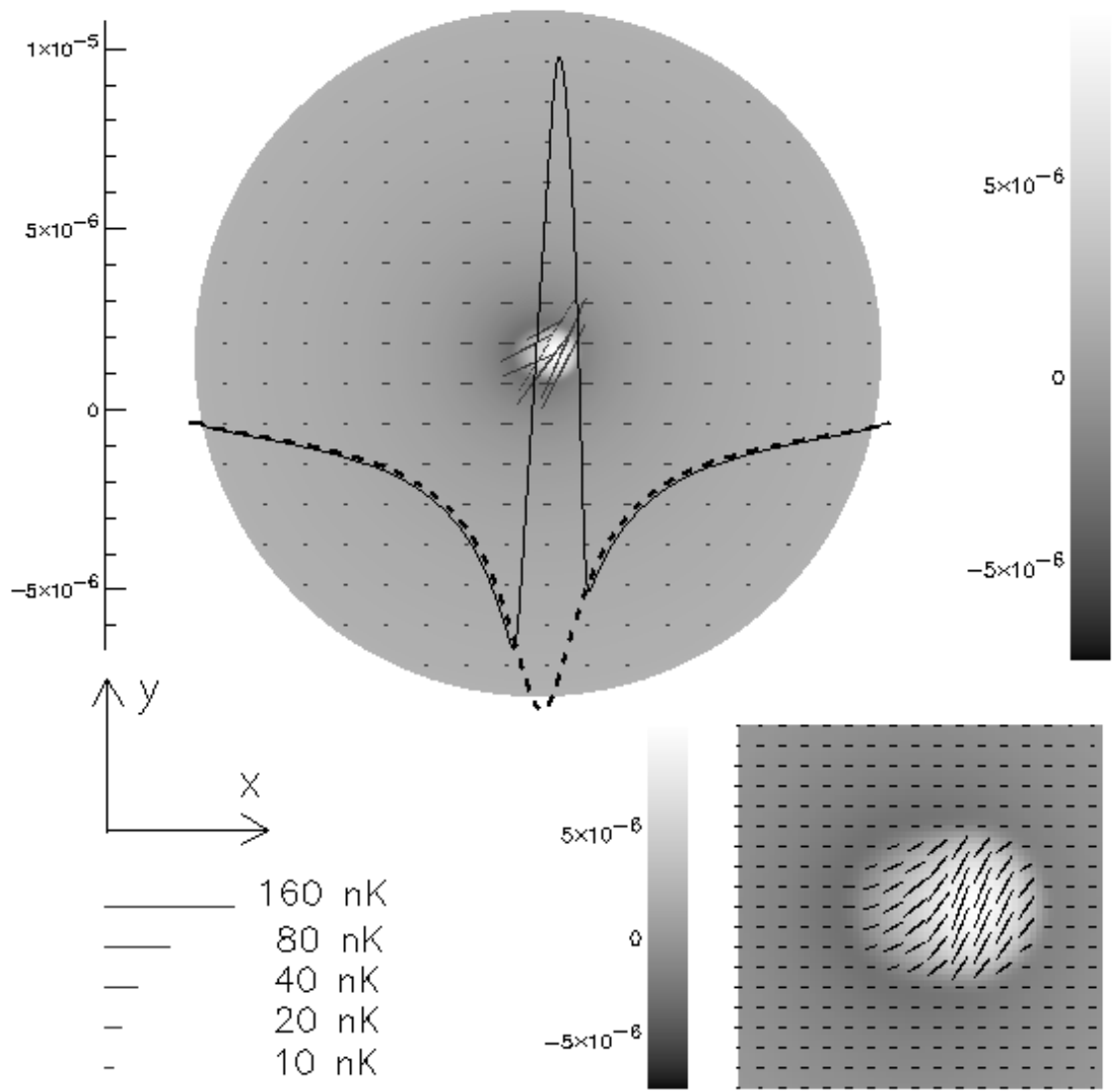}
\caption{
As in figure \ref{fig1} but when the cluster has 
negative velocity along the line of 
sight ($\rm{v}_1 < 0$). The amplitude of the polarization 
is smaller in this case because 
the negative velocity of the cluster partially compensates 
the positive velocity of the central region. The maximum polarization 
is $\approx 160$ nK (300 GHz).
}
\label{fig2}
\end{figure}
\subsection{A  powerful new galaxy cluster investigation tool}
One interesting aspect of the kinetic SZ effect and the CMB polarization 
is that while the first depends on the radial component of the 
velocity, the latter  depends on the tangential component.
Obviously the two measurements are complementary  and can in principle
allow us  to fully reconstruct the 3D velocity vector.
However, the signal we observe is a combination
of the signal associated with the moving gas cloud and the cluster itself.
Thus, the gas cloud velocity 3D reconstruction 
depends on the actual cluster and gas cloud  density profile. 
A steeper profile will show 
more {\it contrast} between the central and 
external parts of the cluster while a shallower profile will show a 
smoother gradient 
in both the kinetic SZ and polarization 2D data. 
Fortunately, information 
about the optical depth can be easily obtained combining the thermal SZ 
effect ($\propto \tau T$) with the temperature information obtained 
from X-rays observations.  
Furthermore, the  X-ray observations of the possible 
cold front associated with a moving cloud will give us 
the absolute velocity of the gas cloud. 
This is a good example of how multi-wavelength observations of the same object 
can be used to reconstruct its 3D structure.
In conclusion, it is clear that, future experiments 
having capabilities to measure the kinetic SZ effect and/or 
polarization of CMB photons induced by galaxy clusters with high
spatial resolution will open new and exciting 
windows for the studies of galaxy cluster dynamics.\\
\acknowledgments
JMD thanks support from a Marie Curie Fellowship of the European 
Community programme 
{\it Improving the Human Research Potential and Socio-Economic knowledge} 
under 
contract number HPMF-CT-2000-00967. This work was
supported by an RTN fellowship, CXC grants GO2-3177X
and the Smithsonian Institution.


\end{document}